\documentclass[aps,prd,twocolumn,superscriptaddress,groupedaddress, showkeys]{revtex4} %,showpacs 
\usepackage{subfig, caption}
\usepackage{natbib}
\usepackage{mathtools}
\usepackage{txfonts}
\usepackage{scalerel}
\usepackage{graphicx}  % needed for figures
\usepackage{bm}        % for math
\usepackage{amssymb}   % for math
\usepackage{lipsum}
\usepackage{lineno}
\usepackage{amsmath,array}
\usepackage{caption}
\usepackage{hyperref}
\hypersetup{
	colorlinks=true,
	linkcolor=blue,
	filecolor=magneta,      
	urlcolor=blue,
}

\DeclareCaptionJustification{justified}{\leftskip=0pt \rightskip=0pt \parfillskip=0pt plus 1fil}

\usepackage{tikz}
\usetikzlibrary{svg.path}
\definecolor{orcidlogocol}{HTML}{A6CE39}
\tikzset{
    orcidlogo/.pic={
        \fill[orcidlogocol] svg{M256,128c0,70.7-57.3,128-128,128C57.3,256,0,198.7,0,128C0,57.3,57.3,0,128,0C198.7,0,256,57.3,256,128z};
        \fill[white] svg{M86.3,186.2H70.9V79.1h15.4v48.4V186.2z}
        svg{M108.9,79.1h41.6c39.6,0,57,28.3,57,53.6c0,27.5-21.5,53.6-56.8,53.6h-41.8V79.1z M124.3,172.4h24.5c34.9,0,42.9-26.5,42.9-39.7c0-21.5-13.7-39.7-43.7-39.7h-23.7V172.4z}
        svg{M88.7,56.8c0,5.5-4.5,10.1-10.1,10.1c-5.6,0-10.1-4.6-10.1-10.1c0-5.6,4.5-10.1,10.1-10.1C84.2,46.7,88.7,51.3,88.7,56.8z};
    }
}
\newcommand\orcidicon[1]{\href{https://orcid.org/#1}{\mbox{\scalerel*{
                \begin{tikzpicture}[yscale=-1,transform shape]
                \pic{orcidlogo};
                \end{tikzpicture}
            }{|}}}}
\begin{document}
	%===============================================
 \title{Bounds on sterile neutrino lifetime and mixing angle with active	neutrinos by global 21 cm signal}
\author{Pravin Kumar Natwariya$^{\orcidicon{0000-0001-9072-8430}}$\,}
\email{pravin@prl.res.in, pvn.sps@gmail.com}
\affiliation{%
    Physical Research Laboratory, Theoretical Physics Division, Ahmedabad, Gujarat 380 009, India}
\affiliation{%
    Department of Physics, Indian Institute of Technology, Gandhinagar, Palaj, Gujarat 382 355, India}  
\author{Alekha C. Nayak$^{\orcidicon{0000-0001-6087-2490}}$\,}
\email{alekhanayak@nitm.ac.in}
\affiliation{%
    National Institute of Technology, Meghalaya, Shillong, Meghalaya 793 003, India}
\date{\today}
%===============================================
\begin{abstract}
Sterile neutrinos can be a possible candidate for dark matter. Sterile neutrinos are radiatively unstable and can inject photon energy into the intergalactic medium (IGM). The injection of photon energy into IGM can modify the temperature and ionization history of IGM gas during cosmic dawn. Theoretical models based on the $\Lambda$CDM framework predict an absorption profile in the 21 cm line during the cosmic dawn era.  Recently, the Experiment to Detect the Global Epoch of Reionization Signature (EDGES) collaboration confirmed such an adsorption signal. Injection of energy into IGM can modify the absorption amplitude in the 21 cm signal. Considering the 21 cm absorption signal at cosmic dawn, we constrain the lifetime of sterile neutrinos and the mixing angle of sterile neutrinos with active neutrinos. We also compare these bounds with other astrophysical observational bounds.
\end{abstract}
%===============================================
%\pacs{}
\maketitle
%===============================================
%===============================================
\section{Introduction}
After about $3\times 10^5$ years of big-bang electron and proton cool down to form the neutral hydrogen atoms. In the $\Lambda $CDM model of cosmology, recombination ends at redshift $z \sim 1010$, and baryonic matter decouples from the cosmic microwave background (CMB) photon. The dark-age begins just after the end of recombination. During the dark-age, the residual free electrons undergo Compton scattering with CMB photons and maintain thermal equilibrium with the gas until $z\sim200$.  After that, gas cools adiabatically, i.e. $T_{\rm gas}\propto(1+z)^2$.  In the $\Lambda $CDM model, gas temperature and free electron fraction between the end of recombination and the cosmic dawn era are well known. The presence of any exotic source of energy can modify the thermal and ionization history of the Universe. Therefore, using observations during the cosmic dawn era, one can put the constraints on such sources of energy injection into IGM.

%=======================================================================

Although the $\Lambda$CDM model of cosmology is highly successful in explaining Big-Bang nucleosynthesis, CMB anisotropies and large length-scale observations such as large scale structure observations, it faces challenges at a smaller length scale, $\lesssim1$~Mpc (for detailed review see \cite{Bullock:2017B} and references therein). These  problems include the missing satellite problem \cite{Moore:1999nt, Klypin:1999uc},  the too-big-to-fail problem \cite{Boylan:2011sm, Boylan:2012bs} and the cusp-core problem \cite{block:2010wj}. In the light of these problems, alternatives to the CDM model have been proposed, for e.g.  self-interacting dark matter (SIDM) \cite{Spergel:2000DSP, Tulin:2017Y, Kaplinghat:2016TY, Natwariya:2020V}, fuzzy cold dark matter \cite{Hu:2000, Schive:2014} or warm dark matter (WDM) \cite{Blumenthal:1982, Dodelson:1994, Colombi:1996, Sitwell:2014, Brdar:2018}. Sterile neutrinos with KeV mass range can be considered as a viable candidate for the WDM (Ref. \cite{Adhikari:2017, Abazajian:2017, Boyarsky:2019} and Refs. therein). Beyond the standard model, the mass generation of active neutrino via see-saw mechanism predicts sterile neutrino. The mass range of the sterile neutrino varies from eV to GUT scale.  As sterile neutrinos are singlet under the standard model gauge group, they can be considered as dark matter. In recent years, different techniques have been proposed to probe the unexplored sterile neutrino DM parameter space \cite{Andrea:2020, Morgan:2020, Brandon:2020, Lopes:2020, Gouva:2020, Seto:2020}. NuSTAR observations did not find any sign of anomalous X-ray lines for sterile neutrino mass range $10-40$~KeV. The future updated version of NuSTAR will be able to probe for sterile neutrino mass range $6-10$~KeV \cite{Brandon:2020}. In the context of EDGES signal, authors of the reference \cite{vipp:2021},  put a constraint on the Dodelson-Widrow sterile neutrinos mass to $63_{-35}^{+19}$~KeV . Further, individual bounds on the sterile neutrino parameter space can be found in the Refs. \cite{Boyarsky:2019I, foster:2021, salvio:2021, Das:2018, Shakeri:2020, Honorez:2017, Vegetti:2018, Rudakovskyi:2018, Bezrukov:2017}. In the present work, we consider radiative decay of sterile neutrino dark matter and study the constraints on its lifetime and mixing angle with active neutrinos. 

%=======================================================================

During the cosmic dawn era, the baryon content of the Universe is dominated by neutral hydrogen and a small fraction of helium. Therefore, the 21 cm signal from the neutral hydrogen atom appears to be a treasure trove to study cosmic history during the cosmic dawn era, test and constraint different type of DM models.  The 21 cm signal arises due to the hyperfine transition between the 1S singlet (0) and triplet (1) states of the neutral hydrogen atom.  The relative number density of the hydrogen atoms in the triplet and singlet states can be parametrized by the spin temperature ($T_S$), i.e. $n_1/n_0= ({g_1}/{g_0})\times\exp[{-E_{10}/T_S}]\,$, where $E_{10}={2 \pi}/{\rm (21~cm)}$.  $g_0$ and $g_1$ are the statistical degeneracies of singlet and triplet states, respectively. In the cosmological scenarios,  the spin temperature is defined by the three competing mechanisms: I) collisions between the Hydrogen atoms or Hydrogen atoms and electrons, II) CMB radiation and III) Ly$\alpha$ radiation from the first stars \cite{Field, Pritchard_2012, Furlanetto2006a},
\begin{alignat}{2}
 T_S^{-1}=\frac{T_R^{-1} + (x_\alpha+x_c)\,T_{\rm gas}^{-1} }{1+x_\alpha+x_c}\,\label{1}\,,
\end{alignat}
$x_c$ is the collisional coupling coefficient while $x_\alpha$ is the Wouthuysen-Field coupling coefficient \cite{1952AJ.....57R..31W, Field, Hirata2006, Mesinger:2011FS}.  For the standard scenarios ($\Lambda$CDM model), one can take $T_R=T_{\rm CMB}$. $T_{\rm CMB}=2.725\,(1+z)$~K is the cosmic microwave background radiation temperature \cite{Fixsen_2009}. The global 21~cm differential brightness temperature is defined as \cite{Furlanetto2006a, Mesinger:2007S, Mesinger:2011FS, Pritchard_2012}, 
\begin{alignat}{2}
T_{21}=27\,x_{\rm HI}\,\left[\frac{0.15}{\Omega_{\rm m }}\,\frac{1+z}{10}\right]^{1/2}\left(\frac{\Omega_{\rm b}h}{0.023}\right)\,  \left(1-\frac{T_{R}}{T_S}\right)~{\rm mK}\,,\label{2}
\end{alignat}
$x_{\rm HI}=n_{\rm HI}/n_H$ is neutral hydrogen fraction in the Universe, $n_{\rm HI}$ and $n_H$ are neutral and total hydrogen number density respectively.   We take the cosmological parameters as: $\Omega_{\rm m }=0.31$, $\Omega_{\rm b}=0.048$ and $h=0.68$  \citep{Planck:2018}. The detectability of $T_{21}$ depends upon the $T_{\rm CMB}/T_S$ ratio. If $T_S >T_{\rm CMB}$, we observe 21 cm emission signal, if  $T_S < T_{\rm CMB}$, one observes 21 cm absorption signal and if  $T_S=T_{\rm CMB}$, there will no any signal. The evolution of the 21~cm signal can be demonstrated using equation \eqref{1}: After the recombination to the redshift $\sim200$, residual electrons undergo Thomson scattering with CMB photons and maintain thermal equilibrium between the gas and the CMB ($T_{\rm gas}=T_{\rm CMB}$). During this period $x_c\gg1$ and $x_\alpha=0$ which implies $T_S=T_{\rm CMB}$ and  no signal is observed. After that, until $z\sim 40$, the gas cools adiabatically, and its temperature falls below CMB. Therefore $T_S < T_{\rm CMB}$ and one observes  early 21~cm absorption signal---known as the collisional absorption signal. Below 50~MHz, radio antennas sensitivity decreases dramatically, and it becomes challenging to observe the collisional absorption signal. Between $z\sim 40$ and first stars formation ($z=z_*$), gas density decreases due to the expansion of the Universe, and collision coupling becomes small ($x_c\ll1$). It implies $T_S = T_{\rm CMB}$, and there will be no any $T_{21}$ signal.  During cosmic dawn ($z\lesssim z_*$),  the Ly$\alpha$ photons from the first stars couple the spin temperature with the gas temperature---Wouthuysen-Field coupling \cite{1952AJ.....57R..31W, Field}.  Therefore, one obtains $T_S=T_{\rm gas}$, and the absorption signal can be observed. Recently, such an absorption signal has been confirmed by the EDGES collaboration. The EDGES observation is centered at 78~MHz or redshift $z = 17.2$ with brightness temperature of $T_{21}^{\rm EDGES} =-500^{+200}_{-500}$~mK with 99 percent confidence limit \cite{Bowman:2018yin}. Considering $T_S=T_{\rm gas}$, the observed brightness temperature translates to gas temperature as $T_{\rm gas}(z=17.2)=3.26^{+1.94}_{-1.58}$~K.  In  the $\Lambda$CDM model, the  gas temperature at redshift  $z=17.2$ remains about 7~K, and corresponds to brightness temperature $T_{21}(z=17.2)\simeq -220$~mK. To resolve the tension between the theoretical prediction based on $\Lambda$CDM model and EDGES observation, one requires to increase the ratio of $T_R/T_S$ over theoretical predictions in redshift range $15\leq z\leq 20$. Either we can increase the background radio radiation or decrease the gas temperature. Both possibilities have been studied by several authors; for example, see the Refs. \cite{Ewall-Wice2018, Biermann:2014, Jana:2018, Feng2018, Lawson:2019,Natwariya:2021, Lawson:2013, Levkov:2020, Fixsen2011, Natwariya:2020, Brandenberger:2019, Chianese:2019,Bhatt2019pac, Tashiro:2014tsa, Barkana:2018lgd, SIKIVIE2019100289, Mirocha2019, Ghara2019}. However, such mechanisms to increase the background radio radiation or cooling the gas are debatable issues. One of such mechanisms to cool gas is baryon dark matter interaction \cite{Barkana:2018lgd}. This approach has been questioned by several authors   \cite{Munoz2018, FRASER2018159, Bransden1958, Barkana:2018nd, Berlin2018, Kovetz2018, Munoz2018a, Slatyer2018}. Here, it is to be noted that the authors do not consider heating of the IGM gas by decaying or annihilating dark matter. Injection of electrons and photons by decaying or annihilating DM into IGM can heat the gas more than cooling of the gas \cite{Amico:2018, Mitridate:2018}.  Subsequently, the EDGES measurement  has been also questioned in several articles. Recently, Shaped Antenna measurement of the background RAdio Spectrum 3 (SARAS 3) observation reported that the  EDGES observation is not an astrophysical origin and it is rejected with the 95.3 percent confidence level \cite{Saurabh:2021}. In the Ref. \cite{Hills:2018}, the authors have questioned the fitting parameters for the foreground emission and data. There is a possibility that the absorption feature in the EDGES observation can be a ground screen artifact \cite{Bradley:2019}. The absorption amplitude may modify depending on modelling of  foreground \cite{Saurabh:2019, Tauscher:2020}. In Ref. \cite{Sims:2019}, the authors perform the Bayesian comparison of fitting models for EDGES data and argue that the highest evidence models favour an amplitude of $|T_{21}|<209$. Recently, the Hydrogen Epoch of Reionization Array (HERA) collaboration reported upper and lower bounds on  the spin temperature of the gas to be: $27{\rm~K}< T_S< 630{\rm~K}\ (2.3{\rm~K}< T_S < 640{\rm~K})$ with confidence level of 68\% (95\%) at redshift $z\sim8$. The corresponding gas temperature found to be: $8.9{\rm~K}< T_{\rm gas}< 1.3\times10^3{\rm~K}$ with 68\% confidence level. However, below the redshift $z\sim10$, the physics of IGM evolution is not very well known. In the light of these constraints and unknown physics at the time of cosmic dawn era, we do not consider any mechanisms to cool IGM gas or increase the background radio radiation. We take 21~cm differential brightness temperature such that it does not change, from its standard value ($\sim-220$~mK), more than a factor of 1/4 or 1/2 at redshift 17.2\,.

%===========================================================================

 \begin{figure*}[htbp!]
 	\begin{center}
 	\subfloat[] {\includegraphics[width=2.35in,height=1.6in]{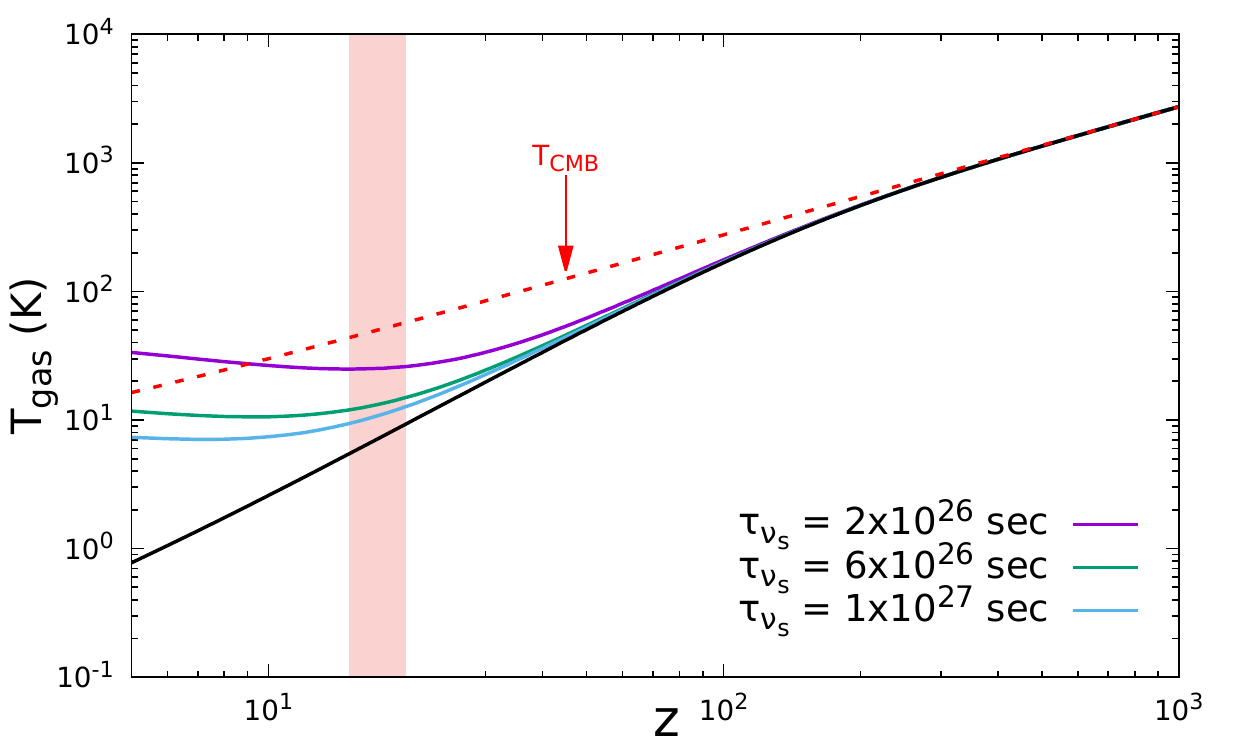}\label{plot:1a}}
 	\subfloat[] {\includegraphics[width=2.35in,height=1.6in]{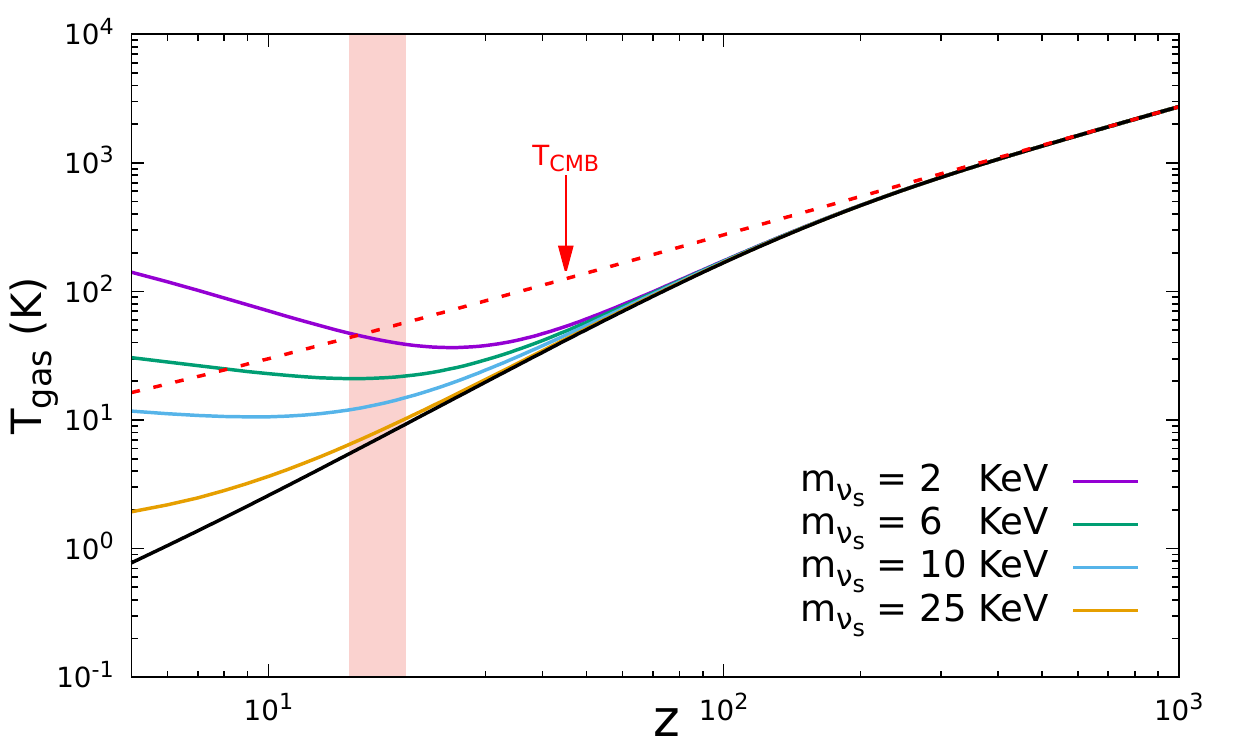}\label{plot:1b}} 
 	\subfloat[] {\includegraphics[width=2.35in,height=1.6in]{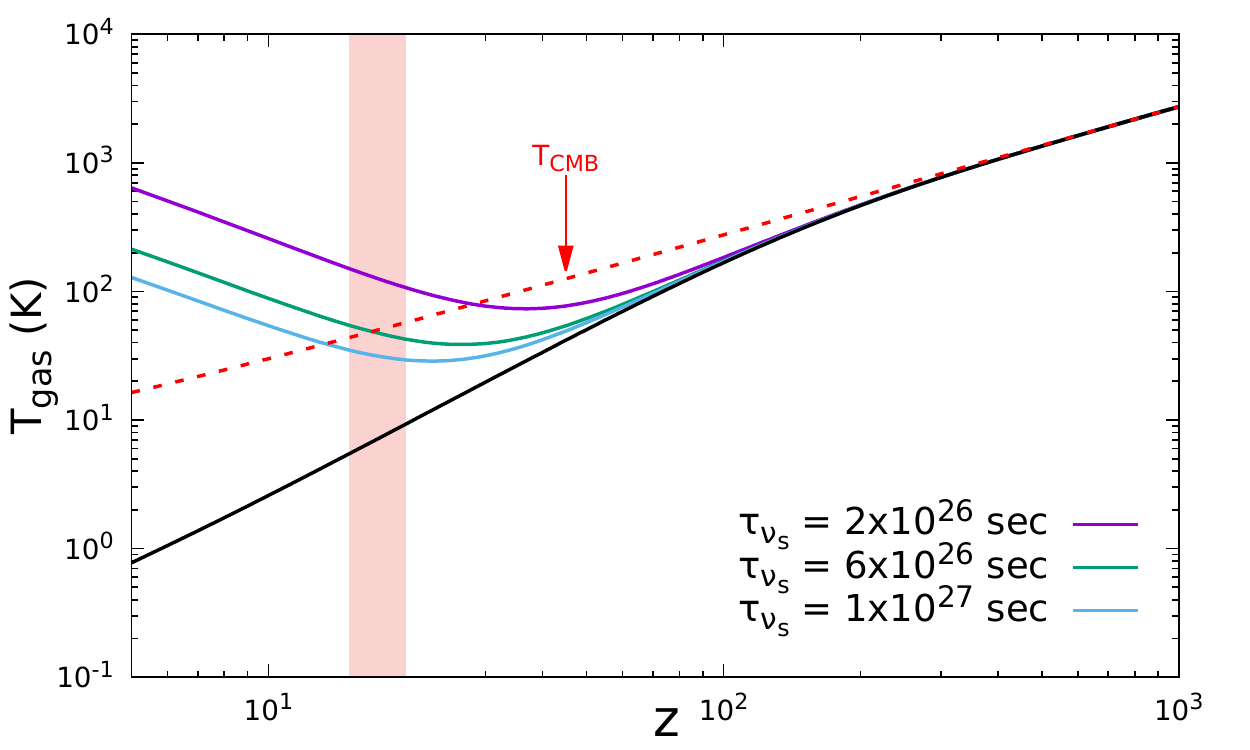}\label{plot:1c}} 
 \end{center}
 \caption{ The gas temperature evolution with redshift in the presence of decaying sterile neutrinos. The red dashed line represents the CMB temperature evolution. The black solid line depicts the $T_{\rm gas}$ when there is no sterile neutrino decay. The shaded region corresponds to EDGES absorption signal, i.e. $15\leq z \leq 20$. In plot \eqref{plot:1a}, we keep mass of sterile neutrino fix to 10~KeV and vary lifetime.  In plot \eqref{plot:1b}, we consider $\tau_{\nu_s}$ constant to $6\times10^{26}$~sec and vary mass of sterile neutrino. In plot \eqref{plot:1c}, we keep $f_{\rm abs}(z,m_{\nu_s})=1/2$ and vary lifetime of sterile neutrino.}\label{plot:1}
\end{figure*}

%===========================================================================

%===========================================================================

\begin{figure*}[htbp!]
	\begin{center}
		\subfloat[] {\includegraphics[width=2.34in,height=1.6in]{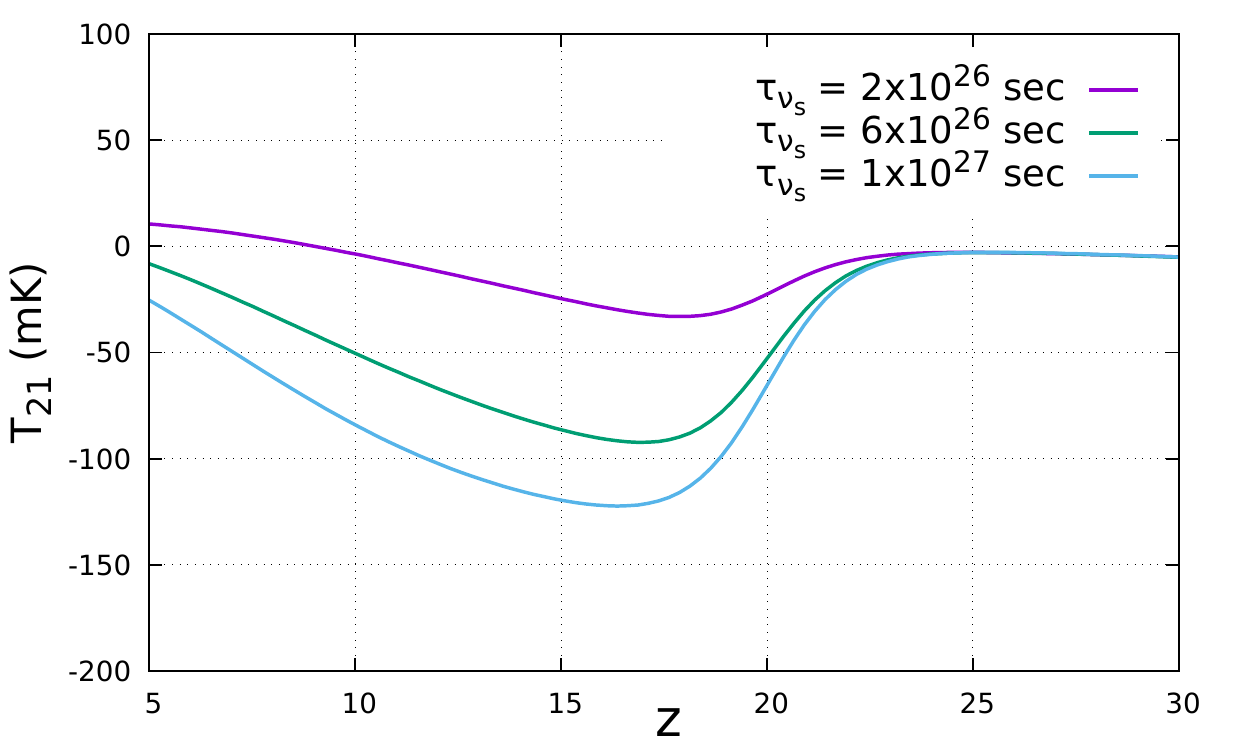}\label{plot:2_1a}}
		\subfloat[] {\includegraphics[width=2.34in,height=1.6in]{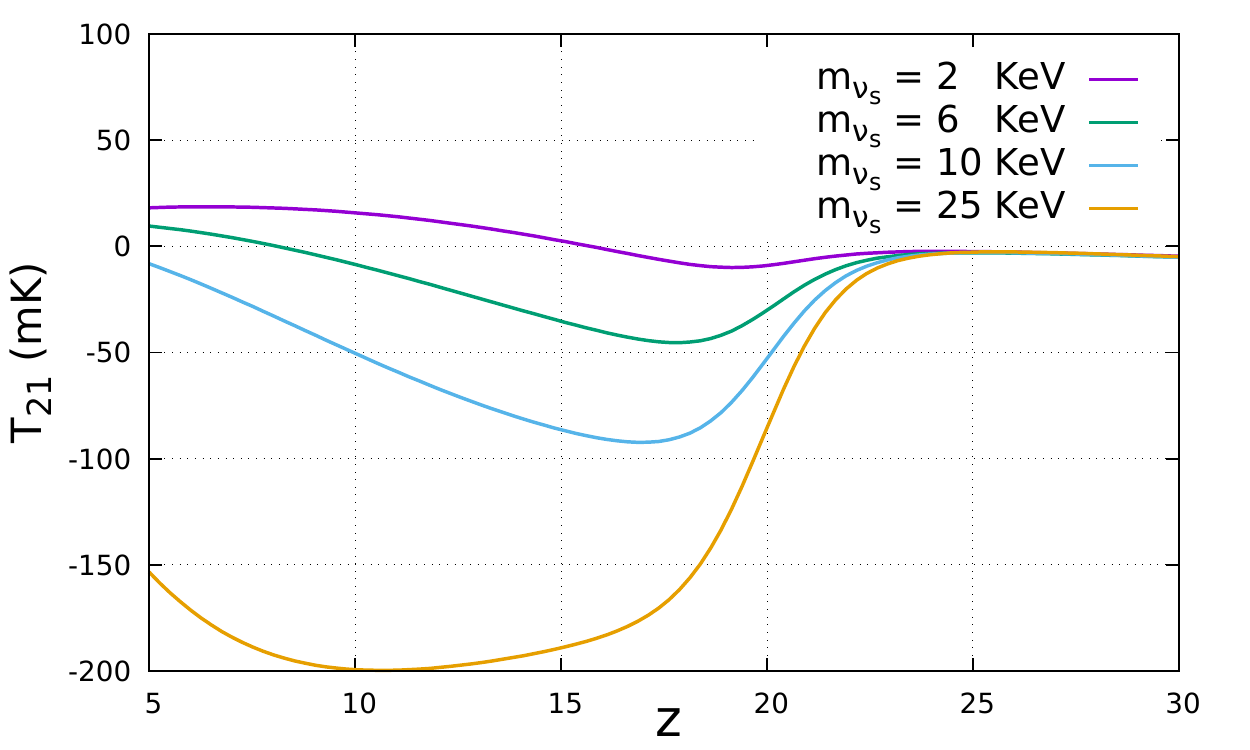}\label{plot:2_1b}} 
		\subfloat[] {\includegraphics[width=2.34in,height=1.6in]{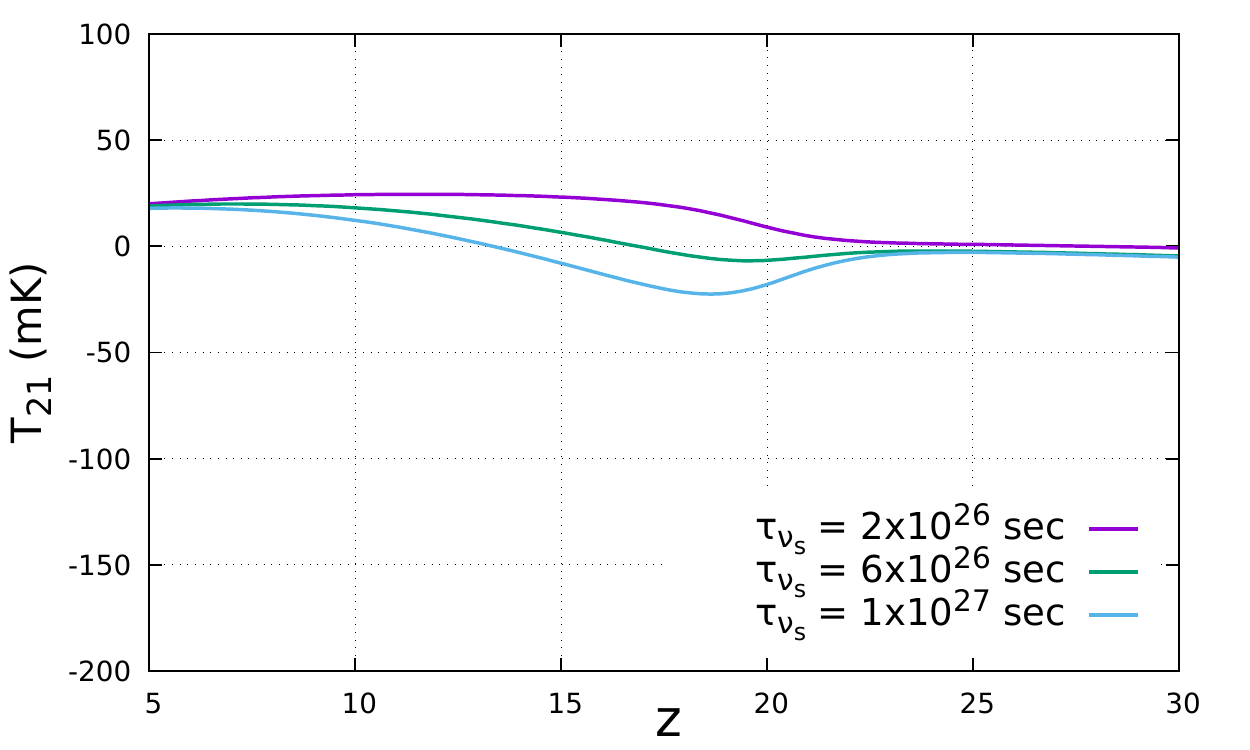}\label{plot:2_1c}} 
	\end{center}
	\caption{ Evolution of 21 cm differential brightness temperature as a function of redshift for the cases represented in figure \eqref{plot:1}.}\label{plot:2_1}
\end{figure*}

%===========================================================================
 
Sterile neutrinos decay to active neutrinos via $\nu_{s}\rightarrow \nu\nu\bar\nu$  and  $\nu_s\rightarrow \nu\, \gamma$ processes. The decay of sterile neutrino dark matter to active neutrino via the radiative process can inject the photon energy into IGM and modify the absorption amplitude of the 21~cm signal during cosmic dawn era. Hence, we can constrain the sterile neutrino decay width and mixing angle with the active neutrino using the 21 cm absorption signal. The decay width of sterile neutrino  for radiative process can be written as  (\cite{Boyarsky:2007, Boyarsky:2019} and Refs. therein),
\begin{alignat}{2}
\Gamma_{\nu_s}=\Gamma_{\nu_s\rightarrow \nu_a \gamma}&=\frac{9\ \alpha\ G_F^2}{1024\,\pi^4}\, \sin^2(2\,\theta)\,m_{\nu_s}^5\nonumber\\
&\simeq 5.52\times 10^{-22}\,\sin^2(\theta)\ \left[\frac{m_{\nu_s}}{\rm KeV}\right]^5\ \left[\frac{1}{\rm sec}\right]\,,\label{3}
\end{alignat}
here, $\theta$ is the mixing angle between sterile and active neutrino, $m_{\nu_s}$ is the mass of the sterile neutrino,  $\alpha$ and $G_F$ are the fine structure and Fermi constant, respectively. The mixing angle $\theta \lll 1$, therefore $\sin^2(2\,\theta)\simeq 4\,\sin^2(\theta)$. The lifetime of sterile neutrino for radiative decay, $\tau_{\nu_s}=1/\Gamma_{\nu_s}\,$.

%==================================================================================

\section{Evolution of the IGM gas and effect of sterile neutrinos}

Evolution of the ionization fraction with redshift in the presence of energy injection by decaying sterile neutrinos \cite{Seager1999, Seager,Liu:2018uzy, Mitridate:2018, Amico:2018, AliHaimoud:2010dx, Galli:2009},
\begin{alignat}{2}
\frac{dx_e}{dz} = \frac{\mathcal{P}}{H\,(1+z)}&\times\,\Big[\,n_H x_e^2\,\alpha_{B}(T_{\rm gas})-(1-x_e)\,\beta_{B}(T_{\rm gas})\,e^{-E_{\alpha}/T_{\rm gas}} \Big]\nonumber\\
&- \frac{1}{H\,(1+z)}\, \Bigg(\,\frac{\mathcal{P}}{E_0}-\frac{1-\mathcal{P}}{E_\alpha}\,\Bigg)\,\frac{(1-x_e)\,\epsilon}{3\,n_H}\,,
\label{4}
\end{alignat}
where $x_e=n_e/n_H$ is the ionization fraction, $n_e$ is the free electron number density in the Universe. $\alpha_{B}$ and $\beta_{B}$ are the case-B recombination coefficient and photo-ionization rate, respectively \cite{Seager1999, Seager, Mitridate:2018}. $E_0=13.6$~eV and $E_\alpha=(3/4)\,E_0$ are ground state binding energy and Ly$\alpha$ transition energy for the hydrogen atom, respectively. $\mathcal{P}$ is the Peebles coefficient \cite{Peebles:1968ja, Mitridate:2018, Amico:2018},
\begin{alignat}{2}
\mathcal{P}=\frac{1+K_H\,\Lambda_H\,n_H\,(1-x_e)}{1+K_H\,(\Lambda_H+\beta_H)\,n_H\,(1-x_e)\,}\,,
\label{5}
\end{alignat}
here, $K_H=\pi^2/(E_\alpha^3\, H)$ and $\Lambda_H=8.22/{\rm sec}$  account for the redshifting of Ly$\alpha$ photon due to expansion of the Universe and the 2S-1S level two photon decay rate of the hydrogen atom, respectively \cite{Tung:1984}. The last term in equation \eqref{4}, describes the additional effect of sterile neutrinos decay on the ionization fraction. $\epsilon\equiv\epsilon(z,m_{\nu_s})$ is the energy deposition rate per unit volume into IGM gas due to decaying sterile neutrinos. It can be written as \cite{Mitridate:2018, Amico:2018, Ripamonti:2006},
\begin{alignat}{2}
\epsilon(z,m_{\nu_s})= {\mathcal{F}}_{S}\,f_{\rm abs}(z,m_{\nu_s})\ \rho_{\nu_s,\rm o}  \ \tau_{\nu_s}^{-1}\ (1+z)^3\label{6}
\end{alignat}
here, $\tau_{\nu_s}$ is the lifetime of sterile neutrino for decay in a active neutrino and a photon. ${\mathcal{F}}_{S}$ is the fraction of the sterile neutrinos that are decaying. We consider that all sterile neutrinos are decaying, i.e. ${\mathcal{F}}_{S}=1\,$.  $\rho_{\nu_s,0}= m_{\nu_s}\,n_{\nu_s,0}$ is the present day energy density of sterile neutrino. $n_{\nu_s,0}$ is the present day number density of sterile neutrinos. For the present work, we consider that all the dark-matter is composed of sterile neutrinos, $\rho_{\nu_s,0}\equiv\rho_{\rm DM,0}\ $, and $\rho_{\rm DM,0}$ is the present day dark-matter energy density \cite{Ripamonti:2006, Mapelli:2005, Dolgov:2002, Boyarsky:2019}. $f_{\rm abs}(z,m_{\nu_s})$ is the energy deposition efficiency into IGM by decaying sterile neutrinos. The energy deposition happens due to only radiative decay of sterile neutrino as active neutrinos interact very weakly with matter. Therefore, we consider only radiative decay of sterile neutrinos. $f_{\rm abs}(z,m_{\nu_s})$ depends on the redshift, mass of sterile neutrino and decay channel \cite{Ripamonti:2006}. The mass of decaying particles enters only through $f_{\rm abs}(z,m_{\nu_s})$. In the presence of energy deposition  into IGM, the gas temperature evolution with redshift \cite{Seager1999, Seager, Liu:2018uzy, Mitridate:2018, Amico:2018, Natwariya:2021PBH, Galli:2009}, 
\begin{alignat}{2}
\frac{dT_{\rm gas}}{dz}  =  \frac{2\,T_{\rm gas}}{(1+z)} +& \frac{\Gamma_{C}}{(1+z)\,H}\, (T_{\rm gas}-T_{\rm CMB}) \nonumber\\
&\ \ -\frac{2}{3\,H\,(1+z)}\times \, \frac{(1+2\,x_e)\ \epsilon}{3\,N_{\rm tot}}%-\frac{\Gamma_{R}}{(1+z)\,(1+f_{He}+x_e)}
\,,
\label{7} 
\end{alignat}
here, $N_{\rm tot}=n_H\,(1+f_{\rm He}+x_e)$ is the total number density of gas, $f_{\rm He}=n_{\rm He}/n_H$ is the Helium fraction and the  Compton scattering rate is defined as,
\begin{equation}
\Gamma_{C}= \frac{8\, \sigma_T\, a_r T_{\rm CMB}^4\, x_e}{3\,(1+f_{\rm He}+x_e)\,m_e}\,,\label{8}
\end{equation}
where, $\sigma_T$, $a_r$ and $m_e$ are the Thomson scattering cross-section, Stefan-Boltzmann radiation constant and mass of electron, respectively. The last term in equation \eqref{7} corresponds to the energy deposition into IGM due to radiative decay of sterile neutrinos.  Following the Refs. \cite{Amico:2018, Mitridate:2018,Chen:2004,Shull:1985}, we consider the `SSCK' approximation--in which $(1-x_e)/3$ fraction of deposited energy goes into ionization, nearly same amount goes into excitation, while $(1+2x_e)/3$ fraction goes into IGM heating. To include the heating of the gas due to energy transfer from CMB photons to the random motions of the gas, we follow the Ref. \cite{Venumadhav:2018} (here we write this heating of the gas as VDKZ18). Authors claim that it can increase the gas temperature by the order of ($\sim 10\%$) at $z\sim17$. Here, it is to be noted that, we do not include the X-ray heating of the gas as it becomes effective after $z\sim17$. Including the heating due to VDKZ18 effect, equation \eqref{7} will modify as,
\begin{alignat}{2}
    \frac{dT_{\rm gas}}{dz}=\frac{dT_{\rm gas}}{dz}\Bigg|_{[{\rm eq. \eqref{7}}]}-\frac{\Gamma_{R}}{(1+z)\,(1+f_{He}+X_e)}\,,\label{9}
\end{alignat}
where, ${dT_{\rm gas}}/{dz}\big|_{[{\rm eq. \eqref{7}}]}$ represents the  temperature evolution in equation \eqref{7}, and heating rate due to energy transfer from CMB photons to the thermal energy of gas by Ly$\alpha$ photons,
\begin{alignat}{2}
    \Gamma_{R}=x_{\rm HI}\,\frac{A_{10}}{2\, H}\,x_{R} \left[\frac{T_R}{T_S}-1\right]\,T_{10}\,,\label{eq13}
\end{alignat}
here, $A_{10}=2.86\times 10^{-15}$~sec$^{-1}$ is the Einstein coefficient for spontaneous-emission from triplet state to singlet state. $x_R=1/\tau_{21}\times[1-\exp(-\tau_{21})]$ and $\tau_{21}=8.1\times10^{-2}\,x_{\rm HI}\,[(1+z)/20]^{1.5}\,(10~{\rm K}/T_S)$ is the 21~cm optical depth. $T_{10}=2\pi\nu_{10}=0.0682$~K. 

%===============================================================================

\section{Results and Discussion}

We solve the coupled equations \eqref{4} and \eqref{7} for different mass and lifetime of sterile neutrino to get $x_{\rm HI}$ and $T_{\rm gas}$ at redshift $z=17.2\,$. To get any absorption signal in redshift range $15-20$, the gas temperature should be less than CMB temperature in shaded region. By requiring $T_{21}\simeq-150$~mK or $-100$~mK at z=17.2, equation \eqref{2}, we can  constraint the lifetime of sterile neutrinos. Subsequently, using equation \eqref{3}, we can also constraint the mixing angle of sterile neutrinos with active neutrinos.
%===============================================================================
\begin{figure*}
    \centering
    \subfloat[] {\includegraphics[width=3.3in,height=2.25in]{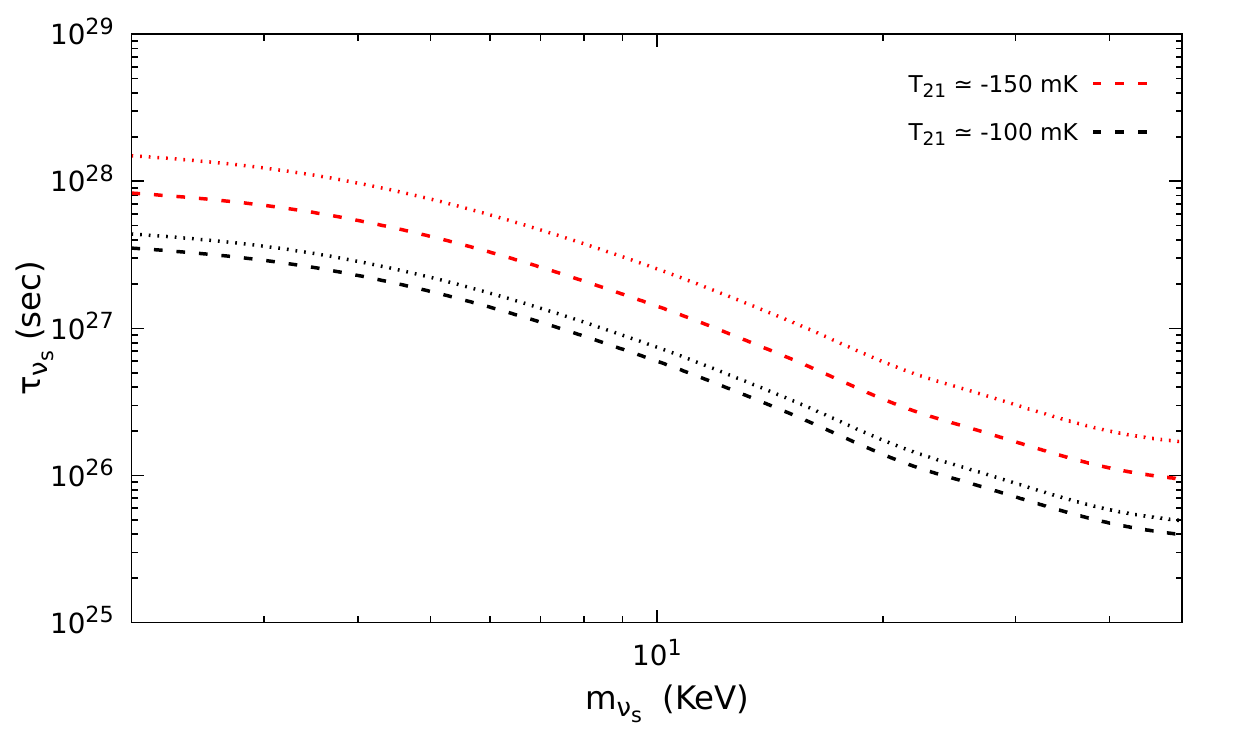}\label{plot:2a}}
    \subfloat[] {\includegraphics[width=3.3in,height=2.25in]{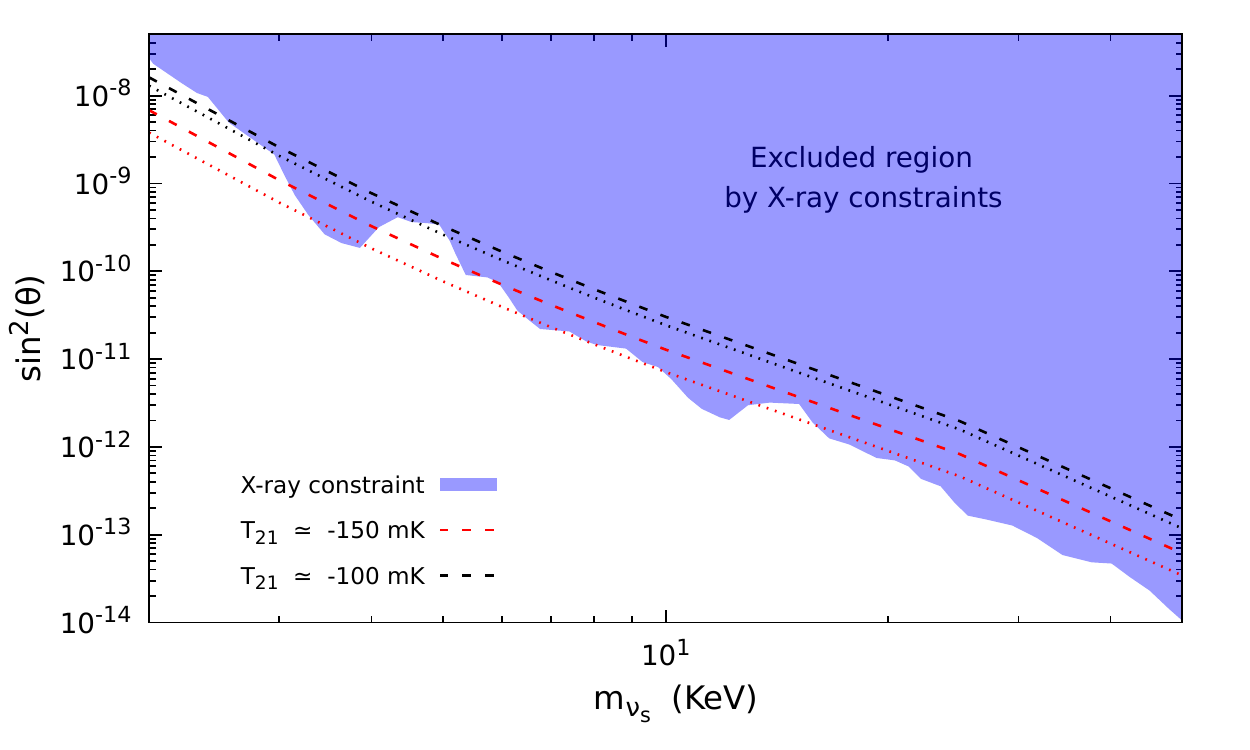}\label{plot:2b}}
    \caption{ In plot \eqref{plot:2a}, we constraint lifetime of sterile neutrinos as a function of mass, while in plot \eqref{plot:2b},  we constraint mixing angle of sterile neutrinos with active neutrinos as a function of mass of sterile neutrinos by keeping 21~cm differential brightness temperature, $T_{21}\simeq-150$ and $-100$~mK. The dotted (dashed) line represents the case when energy transfer from CMB photons to gas is included (excluded) \cite{Venumadhav:2018}. The X-ray constraint on mixing angle (blue shaded region) has been taken from the Ref. \cite{ Boyarsky:2019}.}
    \label{plot:2}
\end{figure*} 
%===============================================================================
In the figure \eqref{plot:1}, we plot the gas temperature evolution as a function of redshift for different mass and lifetime of sterile neutrino. The red dashed line in all plots represents the CMB temperature evolution with redshift. The black solid line represents the gas temperature evolution when their is no effect of decaying sterile neutrino on the IGM gas. The shaded pink region corresponds to redshift range $15\leq z \leq 20\,$. We obtain these results by considering $f_{\rm abs}(z,m_{\nu_s})$ from Ref. \cite{Ripamonti:2006}. In plot \eqref{plot:1a}, we plot the gas temperature for different lifetime of sterile neutrino ($\tau_{\nu_s}$) by keeping the mass ($m_{\nu_s}$) fix to 10~KeV. The violet solid line depicts the gas temperature evolution when lifetime of sterile neutrino is $2\times10^{26}$~sec. As we increase the lifetime of sterile neutrino, the gas temperature decreases–shown by green and cyan curves. It happens because by increasing the $\tau_{\nu_s}$, the radiative decay of sterile
neutrinos decreases and the number of photons injected into
IGM also decreases. Therefore, we get less heating of IGM
by increasing the $\tau_{\nu_s}$. In plot \eqref{plot:1b}, life time of sterile neutrino is fixed to $6\times10^{26}$~sec and the values of $m_{\nu_s}$ varies from $2$~KeV (violet solid line) to $25$~KeV (yellow solid line). If one increases the sterile neutrino mass from 2~KeV (violet line) to 6~KeV (green line), the heating of IGM decreases significantly in the shaded region. It happens because $\rho_{\nu_s}=m_{\nu_s}n_{\nu_s}$, $n_{\nu_s}$ is the number density of sterile neutrinos. Therefore at a particular redshift, when one increases $m_{\nu_s}$ the number density of sterile neutrino decreases, and we get less photon injunction, produced from decaying sterile neutrinos, into the IGM. Hence, one gets less heating of IGM when the mass of sterile neutrino increases. If one considers the immediate and complete absorption of the photon energy into IGM,  then energy deposition efficiency, $f_{\rm abs}=1/2$ ---half of the energy of sterile neutrino will be carried away by active neutrino \cite{Ripamonti:2006, Mapelli:2006}. Mass of the sterile neutrino in the equations \eqref{4} and \eqref{7}, enters through only $f_{\rm abs}$. Therefore, the energy deposition rate, equation \eqref{6}, will depend only on  the lifetime of sterile neutrinos. This case has been depicted in figure \eqref{plot:1c} for the different values of $\tau_{\nu_s}$. In this case, as expected, the heating of IGM increases more compared to the cases in figure \eqref{plot:1a}. 

%=============================================================================

 In figure \eqref{plot:2_1}, we plot the evolution of the 21 cm differential brightness temperature as a function of redshift for the scenarios discussed in figure \eqref{plot:1}.  We consider the $tanh$ parametrization model for the  Wouthuysen-Field coupling coefficient \cite{Kovetz2018, Mirocha:2015G, Harker:2015M}. In the shaded region in figure \eqref{plot:1}, the spin temperature can be approximated as gas temperature. Therefore, when the gas temperature is lower than CMB temperature, we get the absorption profile, i.e. $T_{21}<0$. When the gas temperature rises above the CMB temperature, $T_{21}$ becomes positive, and we see an emission profile. In plot \eqref{plot:2_1}, above the redshift $z\simeq25$, $x_\alpha,\,x_c<1$, therefore, the spin temperature is dominated by CMB temperature, i.e. $T_{21}\approx0$. To get the absorption profile at $z\sim17$, one has to keep $T_{\rm gas}<T_{\rm CMB}$.

%==============================================================================
In figure \eqref{plot:2a}, we plot the lower constraint on lifetime as a function of $m_{\nu_s}$ by requiring  $T_{21}$ such that it does not suppress the standard theoretical value of $T_{21}(z=17.2)\approx-220$~mK more than a factor of 1/4 or 1/2. Considering $T_{21}\lesssim-220$~mK, will further strengthen our bounds. The red coloured curves depict the lower constraint on $\tau_{\nu_s}$ when $T_{21}\simeq-150$~mK, while the black coloured curves represent the lower constraint on $\tau_{\nu_s}$ when $T_{21}\simeq-100$~mK. To get the dashed line, we do not take into account the VDKZ18 heating of the gas. For the dotted line we consider  VDKZ18 heating of the gas. Inclusion of VDKZ18, gives more stringent constraint on $\tau_{\nu_s}$ as gas temperature rises due to the energy transfer from CMB photons mediated by Ly$\alpha$ photons. In figure \eqref{plot:2b}, we obtained the  upper constraint on mixing angle of sterile neutrinos with active neutrinos as a function of mass. For reference, we have also plotted the X-ray constraints on mixing angle as function of $m_{\nu_s}$. The constraints are obtained by assuming solely radiative decay of sterile neutrinos. X-ray constraint comes from the fact that no such X-rays have been seen in observations \cite{Boyarsky:2019}. The red and black coloured curves depict the upper constraint on mixing angle when $T_{21}\simeq-150$~mK and -100~mK, respectively. To get the dashed curves, we do not take into account the VDKZ18 heating of the gas. For the dotted line we have included the VDKZ18 heating of the gas. Here, it is to noted that these bounds do not depend on dark-matter clustering. Therefore, the bounds are free of astrophysical parameters such as density profile or mass function of dark-matter halos, etc. To obtain these bounds, we do not consider any non-standard cooling mechanism to cool the IGM or any source of radio photons. The results in figure \eqref{plot:2}, are comparable with the direct/indirect probe of sterile neutrino dark-matter decay.

%===============================================================================

\section{Summary} 
We have constrained the sterile neutrino dark matter lifetime and mixing angle with active neutrino as a function of sterile neutrino mass. Here, we do not consider the absorption amplitude in the 21 cm line reported by the EDGES collaboration. As discussed above, the EDGES signal has been questioned in several articles. In the light of these controversies over the  EDGES signal, we consider the absorption amplitude predicted by theoretical models based on the $\Lambda$CDM framework of cosmology. We get the bounds such that energy injection from radiative decay of sterile neutrino does not change this absorption amplitude, from its standard value ($\sim-220$~mK), more than a factor of 1/4 or 1/2 at the redshift, $z= 17.2\,$. We have considered the two scenarios to get the bounds: First, IGM evolution without the heat transfer from the background radiation to gas mediated by Ly$\alpha$ photons (VDKZ18 effect). Next, we have considered the VDKZ18 effect on the IGM gas. The following summarises our results for $T_{21} = -150$~mK. 

In the first scenario, the lower bound on the sterile neutrino lifetime varies from $8.3\times10^{27}$~sec to $9.4\times10^{25}$~sec by varying sterile neutrino mass from 2 KeV to 50 KeV.  The lifetime of sterile neutrino decrease when one increases the mass of the sterile neutrino.  It happens because $\rho_{\nu_s}=m_{\nu_s}n_{\nu_s}$. At a particular redshift, when one increases $m_{\nu_s}$, the $n_{\nu_s}$ decreases. Consecutively, one gets less radiative decay of sterile neutrinos. Therefore, we get more window to increase the gas temperature, i.e. we can decrease the lifetime of sterile neutrinos. The upper bound on the mixing angle ($\sin^2\theta$) varies from  $6.8\times10^{-9}$ to  $6.1\times10^{-14}$ by varying sterile neutrino mass from 2 KeV to 50 KeV.

In the second scenario, the lower bound on the sterile neutrino lifetime varies from $1.5\times10^{28}$ sec to  $1.7\times 10^{26}$ sec by varying sterile neutrino mass from 2 KeV to 50 KeV. While the upper bound on the mixing angle varies from $3.8\times10^{-9}$ to $3.42\times10^{-14}$ by varying sterile neutrino mass from 2 KeV to 50 KeV. 

We have also plotted the X-ray constraint to rule out some parameter space for mixing angle of the sterile neutrinos with active neutrinos \cite{Boyarsky:2019}. Although we have considered that sterile neutrinos account for all the dark matter in the Universe, sterile-neutrino may account for only a fraction of the dark matter abundance. In this scenario, the bounds on the sterile neutrino lifetime and mixing angle with active neutrino may modify.

%==============================================================================

\section*{ACKNOWLEDGEMENTS}
The authors would like to thank anonymous referee for the suggestions and a detailed report that significantly improved the quality of the manuscript. A. C. N. is supported by the Science and Engineering Research Board, Republic of India/IN under Grant No. SRG/2021/002291.

\end{document}